\documentclass[10pt,twocolumn,letterpaper]{IEEEtran}
\usepackage{amsmath,epsfig}
\usepackage{algorithm}
\usepackage{bm}
\usepackage{tikz}
\usepackage{graphicx}

\usetikzlibrary{positioning}
\usetikzlibrary{shapes.multipart}
\usepackage{pgfplots}
\usepackage[usenames,dvipsnames]{pstricks}
\usepackage{epsfig}
\usepackage{pst-grad} 
\usepackage{pst-plot} 
\usepackage{subfigure}
\usepackage[capitalize]{cleveref}

\usepackage{subfigure}
\usepackage[T1]{fontenc}
\usepackage{amsmath}
\usepackage{graphics}
\include{psfig}
\usepackage{epsfig}
\usepackage{array}
\usepackage{psfrag}
\usepackage{amssymb}
\usepackage{color}
\usepackage{pstricks,pst-node,pst-text,pst-3d,pst-plot}
\usepackage{ulem}
\definecolor{dgreen}{rgb}{0, 0.8, 0.1}

\def\bw{{\bf w}}

\def\bA{{\bf A}}

\def\bs{{\bf s}}

\usepackage{fancyhdr}
\usepackage{cite}

\makeatletter

\begin{document}

\title{3D Near-Field Virtual MIMO-SAR Imaging Using FMCW Radar Systems at 77 GHz}
\author{\IEEEauthorblockN{Shahrokh Hamidi}\\
\IEEEauthorblockA{Department of Electrical and Computer Engineering, University of Waterloo\\
Waterloo, Ontario, Canada\\
Email: shahrokh.hamidi@uwaterloo.ca}
}

\maketitle

\begin{tikzpicture}[remember picture, overlay]
      \node[font=\small] at ([yshift=-1cm]current page.north)  {This paper has been accepted for publication in the 2024 IEEE Canadian Conference on Electrical and Computer Engineering. \copyright IEEE};
\end{tikzpicture}

\begin{abstract}
In this paper, we present 3D high resolution radar imaging at millimeter-Wave (mmWave) frequencies by means of a combination of virtual Multiple Input Multiple Output (MIMO) Frequency Modulated Continuous Wave (FMCW) Radars and Synthetic Aperture Radar (SAR)  which results in a compact, low-cost, and high-speed 3D mmWave imagery system with low complexity.
\end{abstract}


\section{introduction}
Virtual Multiple Input Multiple Output (MIMO) Frequency Modulated Continuous Wave (FMCW) radar systems operating at mmWave frequencies have recently become popular since they are cost-effective and compact. In addition, they generate signals with high bandwidth which in turn facilitate the possibility of achieving high resolution limits in the range direction \cite{Shahrokh_TDM, Shahrokh_CDM, 24GHzTDM, 77GHzTDM, FDMMIMO}.
The range resolution depends on the bandwidth of the transmitted signal, also known as chirp signal \cite{Skolnik, Mahafza}. However, the angular resolution depends on the effective length of the array relative to the wavelength of the transmitted signal. Therefore, to achieve high angular resolution,  arrays with sizable number of elements are required.
Nevertheless, increasing the effective length of the array comes at considerable cost. Antenna coupling, power consumption, heat-sink issue, complexity of the system, and cost are among the challenges that we encounter if we decide to utilize arrays with sizable number of antennas.
To cope with this issues, the arrays can be generated synthetically by creating relative motion between the radar and the target to be imaged. This is indeed the idea behind Synthetic Aperture Radar (SAR) imaging which has been utilized for decades to produce high resolution images specifically in the field of space-borne as well as air-borne SAR imaging \cite{Cumming, Soumekh, Curlander, Munson_Stripmap, Harger}.
Although creating synthetic aperture is a promising idea for achieving high angular resolution, the SAR systems still suffer from issues such as spending a great amount of time to collect the data over the synthetic aperture.
Virtual MIMO radars are another option to increase the effective length of the array using limited number of TX and RX antennas by exploiting techniques such as Time Division Multiplexing (TDM), Code Division Multiplexing (CDM), and Frequency Division Multiplexing (FDM) methods \cite{Shahrokh_TDM, Shahrokh_CDM, 24GHzTDM, 77GHzTDM, FDMMIMO}.
In fact,  virtual MIMO arrays have recently found their applications in the field of radar imaging and are considered as promising candidates for creating high angular resolution for radar applications \cite{MMW_SAR_Zhuge, Yanik_2D, Yanik_2D_SAR_MIMO, Yanik_3D_SAR_MIMO, Shahrokh_SAR}.
A combination of SAR and virtual MIMO FMCW radars are, therefore, a perfect candidate for high resolution radar imaging systems since they provide imagery systems with lower complexity. Moreover, the amount of time to be spent on the data collection is considerably shorter.

In this paper, we present a 3D high resolution mmWave imagery system by combination of virtual MIMO array processing and the SAR concept using commercially available FMCW radar systems.
In the horizontal direction, a large aperture is created synthetically to produce high angular resolution.
In the vertical direction, however, which using the SAR concept increases the time we need to spend on the data collection process, we will create a large aperture using a sparse MIMO array.
Upon exploiting the virtual MIMO array in the vertical direction and taking advantage of the TDM method, we can then increase the number of elements virtually and this way we can create a large aperture which by combining this idea with the SAR technique, we can then create the effect of a 2D aperture of considerable size  to generate high resolution 3D radar images.
The virtual MIMO FMCW radar operated in this paper, has been designed by Texas Instruments (TI) Inc. It includes 4 similar FMCW radar chips which have been cascaded. Each chip consists of 3 TX and 4 RX antennas. After synchronization, we achieve a system with 12 TX and 16 RX antennas which upon exploiting the TDM method, we can obtain an equivalent system of 1 TX and 192 RX antennas out of which 86 non-overlapped antennas can be selected for the radar imaging application.
The radar system operates at $77 \; \rm GHz$ and is capable of generating approximately $4 \; \rm GHz$ bandwidth.   

The paper has been organized as follows. In Section \ref{System Model}, we address the system model and formulate the imaging problem. In Section \ref{Image Reconstruction}, we present the procedure required to process the raw data and reconstruct the image. Finally, Section \ref{Experimental Results} has been dedicated to the experimental results based on the data gathered from a cascaded-chip FMCW radar operating at $77 \; \rm GHz$ followed by the concluding remarks.
\section{System Model}\label{System Model}
\begin{figure}[htb]
\centering
\begin{tikzpicture}[yshift=0.00001cm][font=\Large]
\node(img1) {\includegraphics[height=5.5cm,width=6cm]{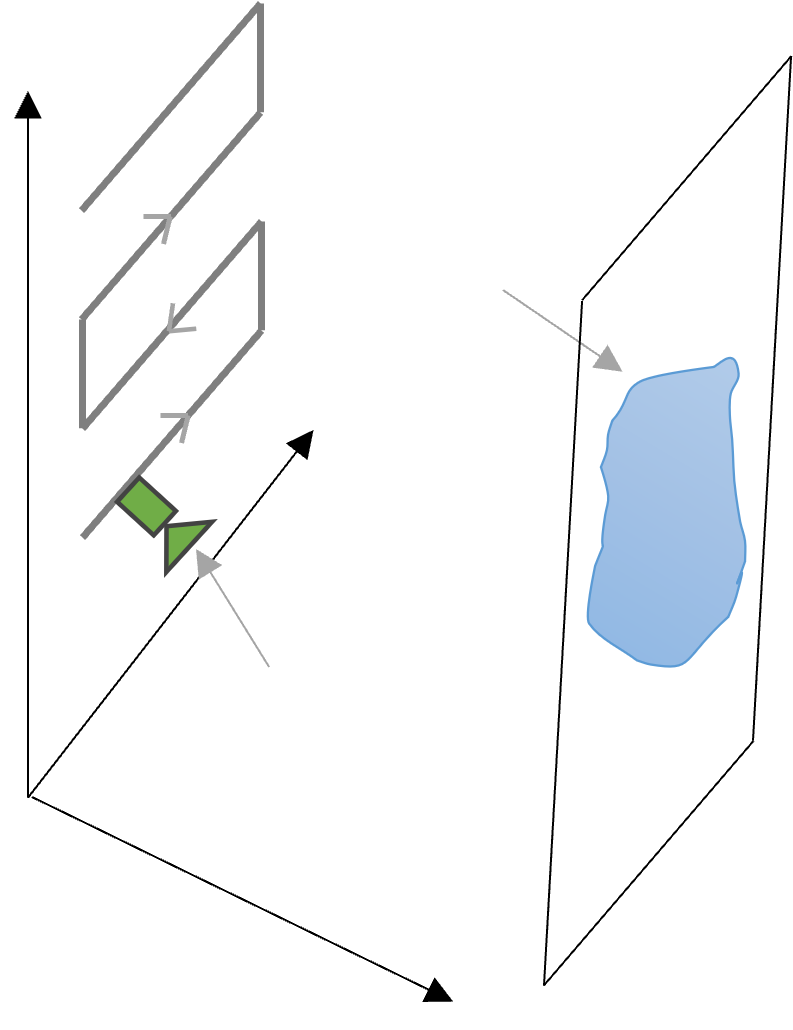}};
 \node[left=of img1, node distance=0cm, xshift=5.5cm, yshift=1.5cm,font=\color{black}] {{Target}};
 \node[left=of img1, node distance=0cm, xshift=4.5cm, yshift=-1cm,font=\color{black}] {{RADAR}};

\end{tikzpicture}
\caption{The geometry of the model.}
\label{fig:Model_Geometry}
\end{figure}

In this section, we describe the system model. Fig.~\ref{fig:Model_Geometry} illustrates the geometry of the imaging problem.
The radar is mounted on a two-axis motorized scanner which spans the x-y plane to create the 2D synthetic aperture. The aperture plane is located at $z=0$. Next, we consider a continuum of point reflectors in $(x,y,z)$ space on the plane situated at $z = z_d$.
When the target is illuminated by the transmitted signal, according to the Huygens principle, each point reflector can be considered as a secondary point source and they, in turn, radiate a portion of the received energy back toward the radar. The signal received at the location of the RX antenna, from a continuum of point reflectors located at $z = z_d$, is described as
\begin{align}
\label{beat_signal_continuum_1}
& s_0(k,x_t,y_t,x_r,y_r,z_d) =  \int \int p(x^{\prime},y^{\prime},z_d)\times \nonumber \\ &e^{\displaystyle -jk\sqrt{(x_t-x^{\prime})^2+(y_t-y^{\prime})^2+z_d^2}}\times \nonumber \\ &e^{\displaystyle -jk\sqrt{(x_r-x^{\prime})^2+(y_r-y^{\prime})^2+z_d^2}}dx^{\prime}dy^{\prime}.
\end{align}
In (\ref{beat_signal_continuum_1}), $k = \frac{2\pi}{c}(f_0+\beta t)$, in which, $\beta = \frac{b}{T}$ where $b$ and $T$ stand for the bandwidth and the chirp time, respectively. The complex parameter $p(x^{\prime},y^{\prime},z_d)$ is the reflection coefficient of the point reflector located at $(x^{\prime},y^{\prime},z_d)$. Furthermore, $f_0$ is the start frequency and $c$ is the speed of light in vacuum. Additionally, $(x_t, y_t)$ and $(x_r, y_r)$ refer to the location of the transmit and receive antennas, respectively.

The goal of radar imaging is to estimate the complex reflective coefficient field $p$. The FMCW radar system we consider in this paper, is a multi-static radar system. This means that the TX and RX antennas are separated spatially. In order to be able to perform the imaging procedure efficiently, however, it is better to have each RX antenna at the same location as its corresponding TX antenna. In other words, to implement the imaging algorithm efficiently, we are required to perform multi-static to mono-static transformation \cite{MMW_SAR_Zhuge, Yanik_2D_SAR_MIMO, Yanik_3D_SAR_MIMO_}. To accomplish this goal, we move the location of the $i^{\rm th}$ TX-RX pair to the midpoint value which upon performing that we can rewrite (\ref{beat_signal_continuum_1}) as
\begin{align}
\label{beat_signal_continuum_2}
& s_0(k,x,y,z_d) =  \int \int p(x^{\prime},y^{\prime},z_d)\times \nonumber \\ &e^{\displaystyle -jk\sqrt{((x + \frac{\Delta_x}{2}) - x^{\prime})^2+((y + \frac{\Delta_y}{2}) -y^{\prime})^2+z_d^2}}\times \nonumber \\ &e^{\displaystyle -jk\sqrt{((x - \frac{\Delta_x}{2}) -x^{\prime})^2+((y - \frac{\Delta_y}{2})-y^{\prime})^2+z_d^2}}dx^{\prime}dy^{\prime},
\end{align}
where $ x = \frac{x_t+x_r}{2}$, $ y = \frac{y_t+y_r}{2}$, $ \Delta x = \frac{x_t-x_r}{2}$ and $ \Delta y = \frac{y_t-y_r}{2}$.
Finally, we can express (\ref{beat_signal_continuum_2}) as
\begin{align}
\label{beat_signal_continuum_3}
& s_0(k,x,y,z_d) = e^{\displaystyle j \Phi (\Delta_x, \Delta_y, z_d)} \int \int p(x^{\prime},y^{\prime},z_d)\times \nonumber \\ &e^{\displaystyle -j2k\sqrt{(x - x^{\prime})^2+(y + y^{\prime})^2+z_d^2}} dx^{\prime}dy^{\prime},
\end{align}
in which $\Phi (\Delta_x, \Delta_y, z_d)$ is an extra phase shift due to the midpoint transformation. After the multi-static to mono-static transformation is preformed, the multi-static radar is transformed into a mono-static radar system.  By bringing the extra phase term $\Phi (\Delta_x, \Delta_y, z_d)$ to the left hand side of the equation (\ref{beat_signal_continuum_3}), we obtain the following
\begin{align}
\label{beat_signal_continuum_4}
& s(k,x,y,z_d) = \int \int p(x^{\prime},y^{\prime},z_d)\times \nonumber \\ &e^{\displaystyle -j2k\sqrt{(x - x^{\prime})^2+(y  -y^{\prime})^2+z_d^2}} dx^{\prime}dy^{\prime}.
\end{align}
In (\ref{beat_signal_continuum_4}), $s(k,x,y,z_d) = e^{\displaystyle -j \Phi (\Delta_x, \Delta_y, z_d)} s_0(k,x,y,z_d)$.
We can further rewrite (\ref{beat_signal_continuum_4}) as
\begin{align}
\label{beat_signal_conv}
s(k,x,y,z_d) =  p(x,y,z_d)\ast_x\ast_ye^{\displaystyle -j2k\sqrt{x^2+y^2+z_d^2}},
\end{align}
where $\ast_x$ and $\ast_y$ represent the convolution operation in the x and y directions, respectively.
From (\ref{beat_signal_conv}), it is clear that $e^{\displaystyle -j2k\sqrt{x^2+y^2+z_d^2}}$ is the 2D impulse response (IPR) of the system.
\section{Image Reconstruction}\label{Image Reconstruction}
Among the well-known techniques for image reconstruction, we can mention the Time Domain Correlation (TDC) method \cite{Soumekh}, the Back Projection (BP) technique\cite{Soumekh}, the Chirp Scaling Algorithm (CSA) \cite{Cumming}, and the wave-number method which is also known as Range Migration Algorithm (RMA) and $\omega-k$ technique \cite{Soumekh, Cumming, MMW_SAR_Zhuge, Cumming_wk}.
The TDC is a highly time consuming technique which is implemented in the fast-time time domain and slow-time time domain.
Although the BP method is implemented in the fast-time frequency domain and slow-time time domain and is more efficient than the TDC technique, it is still computationally prohibitive.
The $\omega-k$ algorithm is based on the 2D convolution formula presented in (\ref{beat_signal_conv}). To perform the 2D convolution efficiently, the algorithm is implemented in the fast-time frequency domain and slow-time frequency domain which by exploiting the Fast Fourier Transform (FFT) technique, it can be implemented effectively \cite{Cumming}.
However, since the multi-static to mono-static transformation is a range dependent process, therefore, we will not implement the traditional $\omega-k$ algorithm which is based on reference phase multiplication and Stolt interpolation \cite{Cumming}. We, instead, take a 3D Fourier transform from (\ref{beat_signal_conv}), which based on the principle of stationary phase (POSP) method \cite{Optics}, is given as follows,
\begin{align}
\label{beat_signal_3dfft}
S_{z_d}(k_x,k_y) = & P(k_x,k_y)\times \nonumber \\
&e^{\displaystyle j\left(z_d \sqrt{4k_c^2 - k^2_x - k^2_y}-x^{\prime}k_x-y^{\prime}k_y \right)},
\end{align}
where $P(k_x,k_y) = \mathfrak{F}_{3D}\{p(x,y,z_d)\}$, in which $\mathfrak{F}_{3D}$ stands for 3D Fourier transform and $S_{z_d}(k_c,k_x,k_y)$ represents the energy of the signal at the range cell corresponding to the plane located at $z = z_d$. Furthermore, $k_c$ represents the wave-number at the center frequency of the transmitted signal.
The next step is to propagate the energy of the signals back to the plane situated at $z = z_d$. This is the plane that the point targets are located. To accomplish this goal, we apply $e^{\displaystyle -j z_{d} \sqrt{4k_c^2 - k^2_x - k^2_y}}$ to the signal presented in  (\ref{beat_signal_3dfft}).
Finally, we perform the inverse 2D Fourier transform to obtain the image,
\begin{align}
\label{RMA_Algorithm_f}
\hat{p}(x,y,z_d) =
  \mathfrak{F}^{-1}_{2D}\left (S_{z_d}(k_x,k_y)e^{\displaystyle - j z_d \sqrt{4k_c^2 - k^2_x - k^2_y}}\right),
\end{align}
where $\mathfrak{F}^{-1}_{2D}$ stands for the 2D inverse Fourier transform in the $x$ and $y$ directions and $\hat{p}(x,y,z_d)$ is the estimated value for the complex reflective coefficient of the targets on the plane that has been located at $z = z_d$.
As can be seen from (\ref{RMA_Algorithm_f}), we first perform phase cancellation per each given value of $z = z_d$. We will then take the 2D inverse Fourier transform in the $x$ and $y$ directions. By choosing different values for $z_d$, we can reconstruct the image of the targets located at different distances from the radar and this way we can obtain a 3D image from the scene to be imaged.
\section{Experimental Results}\label{Experimental Results}
\begin{figure}[htb]
\centering
\begin{tikzpicture}[yshift=0.00001cm][font=\small]
\node(img1) {\includegraphics[height=4.5cm,width=7cm]{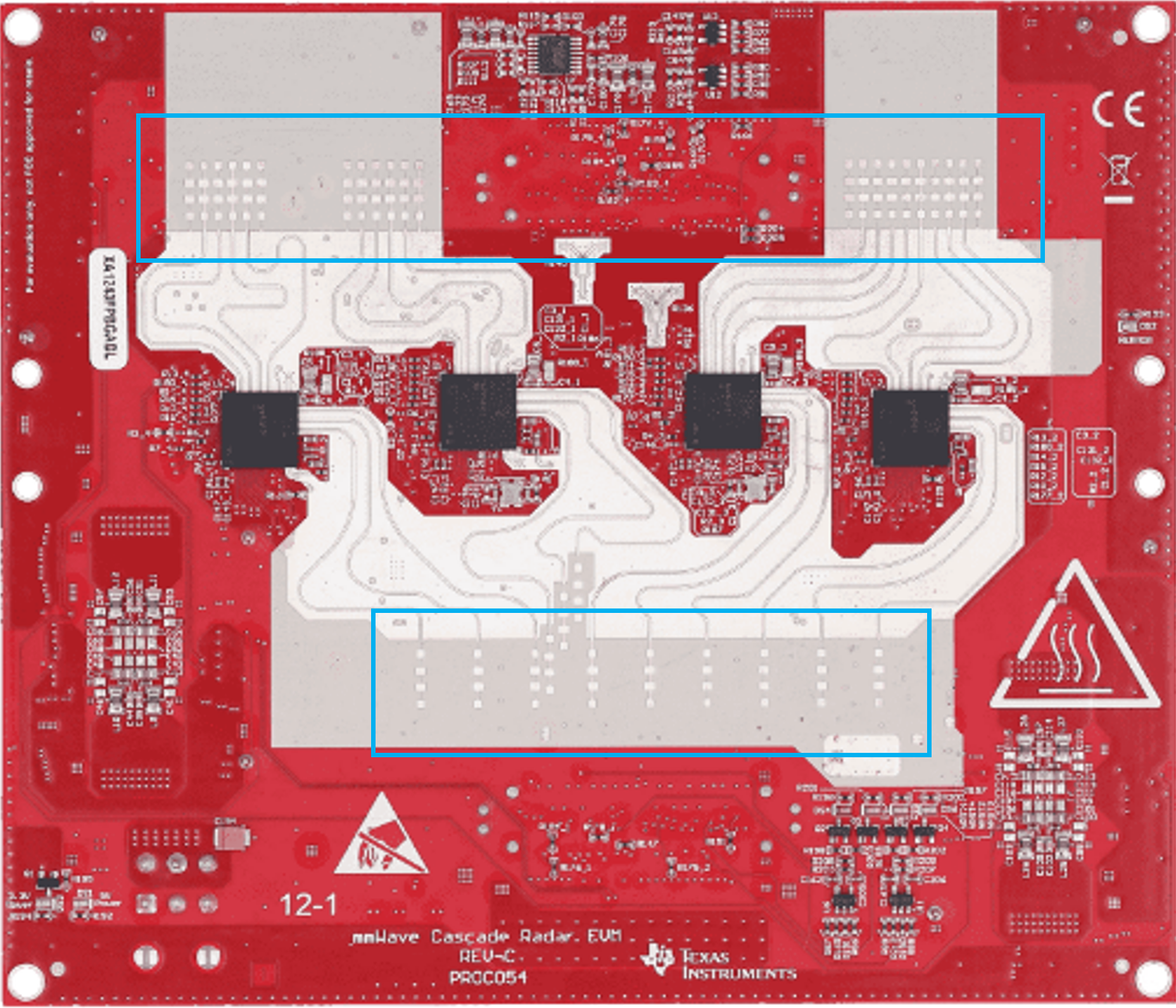}};
 \node[left=of img1, node distance=0cm, xshift=4.5cm, yshift=1.9cm,font=\color{black}] {{Receive Antennas}};
 \node[left=of img1, node distance=0cm, xshift=6.2cm, yshift=-0.35cm,font=\color{black}] {{Transmit Antennas}};

\end{tikzpicture}
\caption{The RF board which consists of 4 synchronized AWR1243 chips. Each chip includes 3 TX and 4 RX antennas.}
\label{fig:CASCADEBoard_1}
\end{figure}

\begin{figure}
\centering
\includegraphics[height=5.5cm,width=8cm]{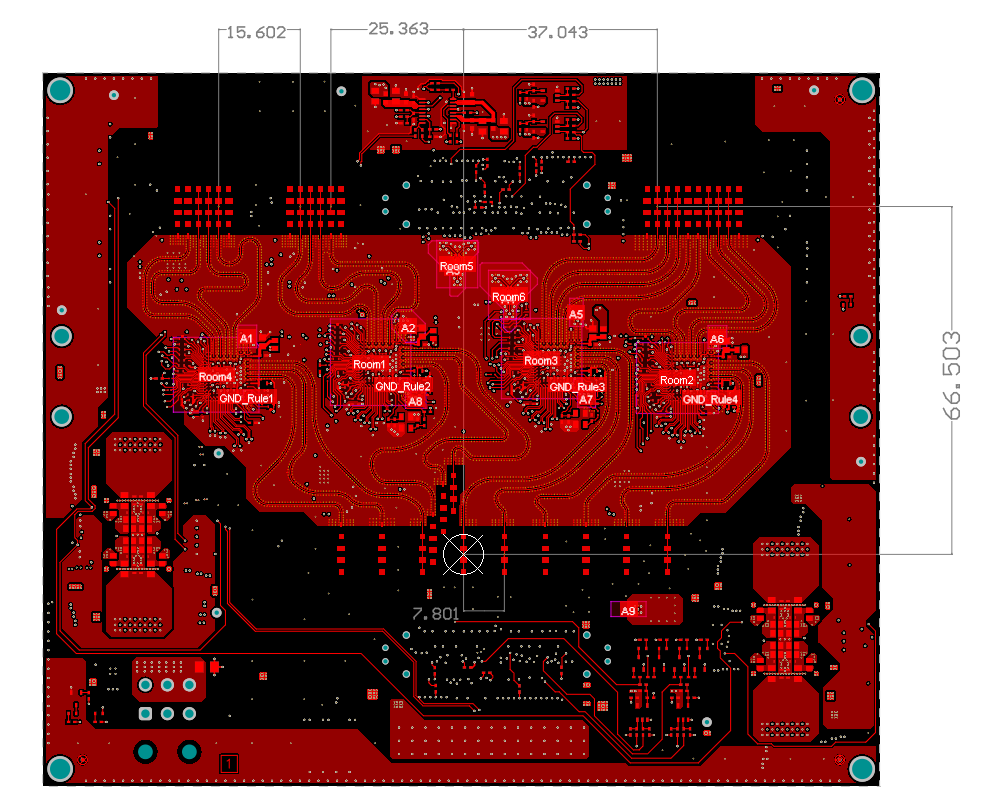}
\caption{The layout of the RF board. The dimensions are in mm.
\label{fig:layout}}
\end{figure}

\begin{figure}[htb]
\centering
\begin{tikzpicture}[yshift=0.00001cm][font=\Large]
\node(img1) {\includegraphics[height=6cm,width=8cm]{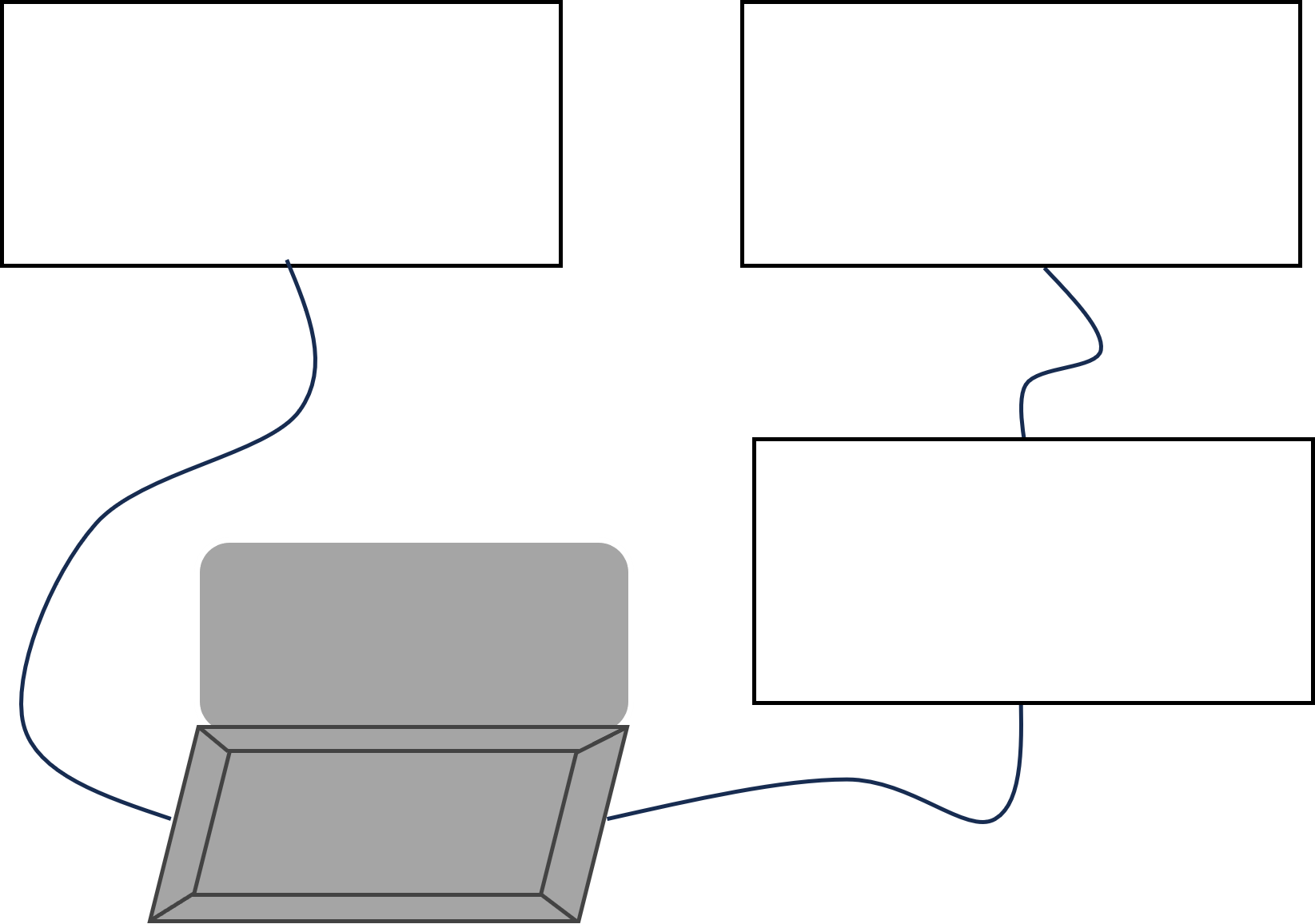}};
 \node[left=of img1, node distance=0cm, xshift=3.5cm, yshift=2.7cm,font=\color{gray}] {{MATLAB}};
 \node[left=of img1, node distance=0cm, xshift=4.7cm, yshift=1.8cm,font=\color{black}] {{MIMO RADAR}};

 \node[left=of img1, node distance=0cm, xshift=7.4cm, yshift=2.7cm,font=\color{gray}] {{Python}};
 \node[left=of img1, node distance=0cm, xshift=9.1cm, yshift=1.8cm,font=\color{black}] {{Stepper Motors}};

 \node[left=of img1, node distance=0cm, xshift=7cm, yshift=-0.1cm,font=\color{gray}] {{C++}};
 \node[left=of img1, node distance=0cm, xshift=8.9cm, yshift=-1.1cm,font=\color{black}] {{Arduino DUE}};

 \node[left=of img1, node distance=0cm, xshift=4.5cm, yshift=-1.2cm,font=\color{black}] {{Laptop}};

\end{tikzpicture}
\caption{A schematic which shows how different parts are connected. The laptop sends commands to the Radar and the stepper motors simultaneously.}
\label{fig:schematic}
\end{figure}

\begin{figure}[htb]
\begin{tikzpicture}[yshift=0.00001cm][font=\tiny]
\node(img1) {\includegraphics[height=2cm,width=3.5cm]{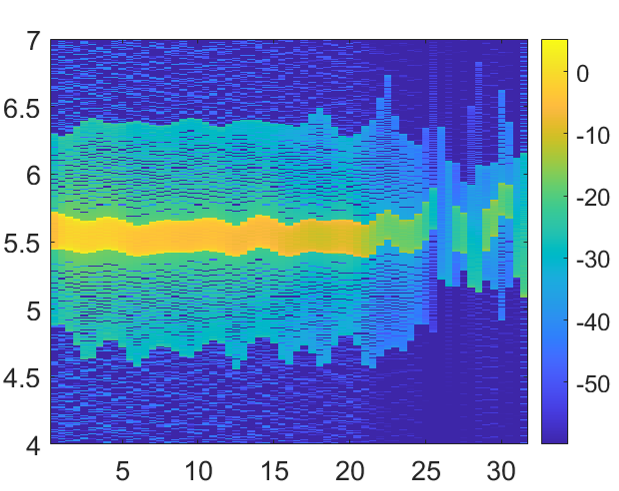}};
 \node[left=of img1, node distance=0cm, rotate = 90, xshift=1cm, yshift=-4.7cm,font=\color{black}] {{Power/Frequency [dB/Hz]}};
 \node[left=of img1, node distance=0cm, rotate = 90, xshift=0.8cm, yshift=-0.9cm,font=\color{black}] {{Frequency [MHz]}};
  \node[below=of img1, node distance=0cm, xshift=0cm, yshift=1cm,font=\color{black}] {{Time [$\mu$s]}};
 \node[below=of img1, node distance=0cm, xshift=0cm, yshift=0.7cm,font=\color{black}] {{(a)}};
\hspace{4.4cm}
\node(img2) {\includegraphics[height=2cm,width=3.5cm]{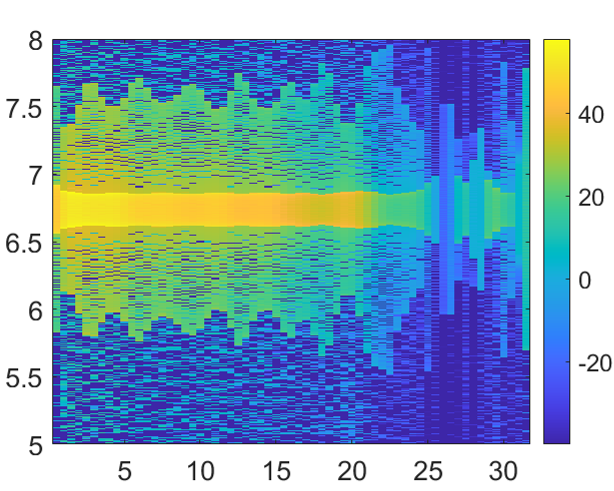}};
 \node[left=of img2, node distance=0cm, rotate = 90, xshift=1cm, yshift=-4.7cm,font=\color{black}] {{Power/Frequency [dB/Hz]}};
 \node[left=of img2, node distance=0cm, rotate = 90, xshift=0.8cm, yshift=-0.9cm,font=\color{black}] {{Frequency [MHz]}};
  \node[below=of img2, node distance=0cm, xshift=0cm, yshift=1cm,font=\color{black}] {{Time [$\mu$s]}};
 \node[below=of img2, node distance=0cm, xshift=0cm, yshift=0.7cm,font=\color{black}] {{(b)}};

\end{tikzpicture}
\caption{a) the spectrogram of the beat signal corresponding to the TX1-RX1 channel from the reference corner reflector which shows the chirp non-linearity, b) the spectrogram of the same signal after compensating for the chirp non-linearity.}
\label{fig:spectrogram}
\end{figure}

\begin{figure}[htb]
\begin{tikzpicture}[yshift=0.00001cm][font=\small]
\node(img1) {\includegraphics[height=2cm,width=3.5cm]{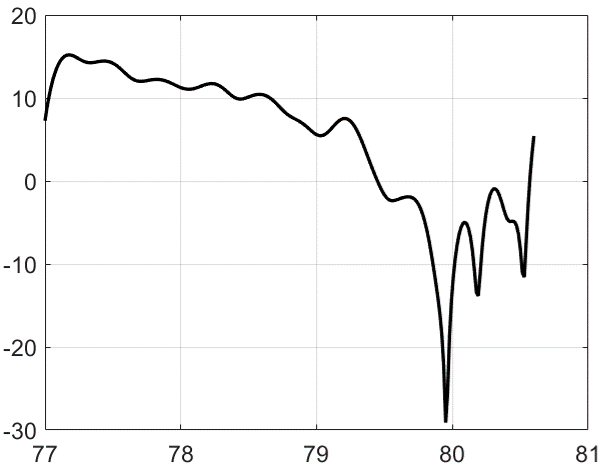}};
 \node[left=of img2, node distance=0cm, rotate = 90, xshift=1.2cm, yshift=-0.9cm,font=\color{black}] {{Power [dBm/Hz]}};
  \node[below=of img2, node distance=0cm, xshift=0.1cm, yshift=1.1cm,font=\color{black}] {{Frequency [GHz]}};
 \node[below=of img2, node distance=0cm, xshift=0cm, yshift=0.7cm,font=\color{black}] {{(a)}};
\hspace{4.4cm}
\node(img2) {\includegraphics[height=2cm,width=3.5cm]{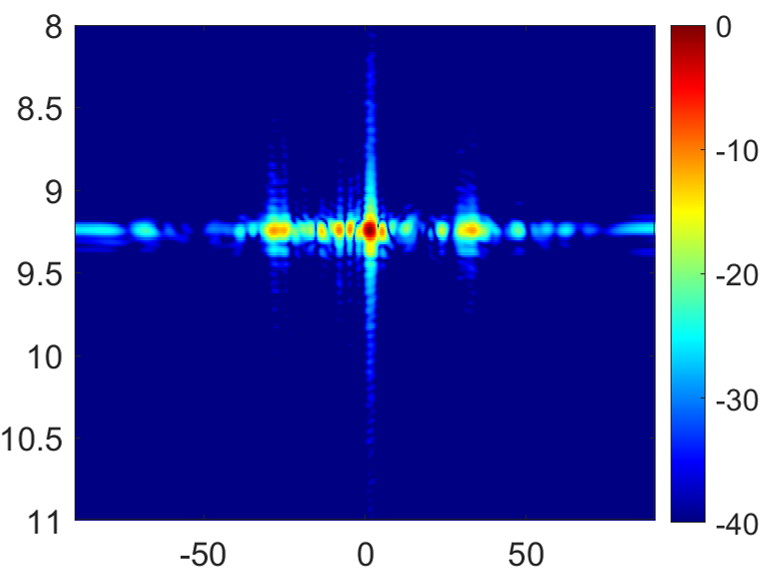}};
 \node[left=of img1, node distance=0cm, rotate = 90, xshift=0.8cm, yshift=-0.9cm,font=\color{black}] {{Range [m]}};
  \node[below=of img1, node distance=0cm, xshift=0cm, yshift=1.1cm,font=\color{black}] {{ $\Theta ^ o$}};
 \node[below=of img1, node distance=0cm, xshift=0cm, yshift=0.7cm,font=\color{black}] {{(b)}};
\end{tikzpicture}
\caption{a) the power spectrum of the beat signal presented in Fig.~\ref{fig:spectrogram}, b) the 2D IPR of the array based on a $20 \;\rm dbsm$ corner reflector located at bore-sight and at $8.95 \;\rm m$ radial distance from the cascaded-chip radar system.}
\label{fig:power_spectrum}
\end{figure}

\begin{figure}[htb]
\begin{tikzpicture}[yshift=0.00001cm][font=\tiny]
\node(img1) {\includegraphics[height=2cm,width=3.5cm]{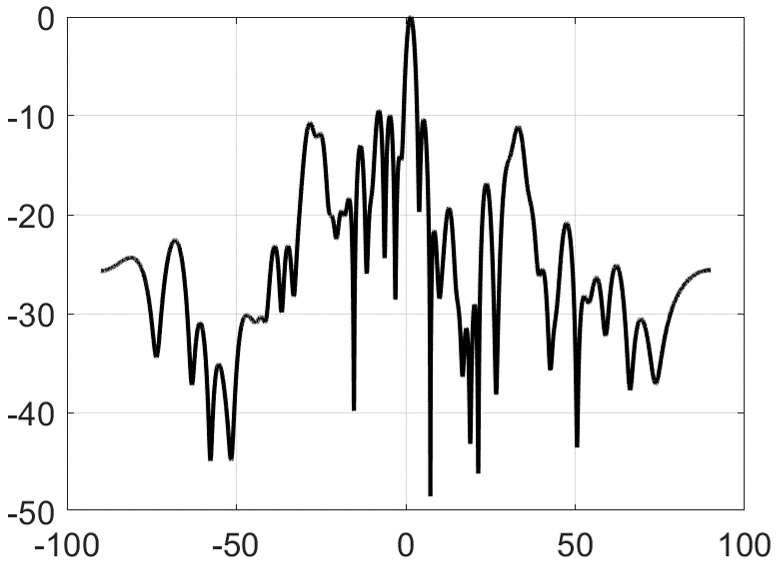}};

 \node[left=of img1, node distance=0cm, rotate = 90, xshift=0.9cm, yshift=-0.9cm,font=\color{black}] {{Normalized Power [dB]}};
  \node[below=of img1, node distance=0cm, xshift=0cm, yshift=1.1cm,font=\color{black}] {{ $\Theta ^ o$}};
 \node[below=of img1, node distance=0cm, xshift=0cm, yshift=0.8cm,font=\color{black}] {{(a)}};
\hspace{4.4cm}
\node(img2) {\includegraphics[height=2cm,width=3.5cm]{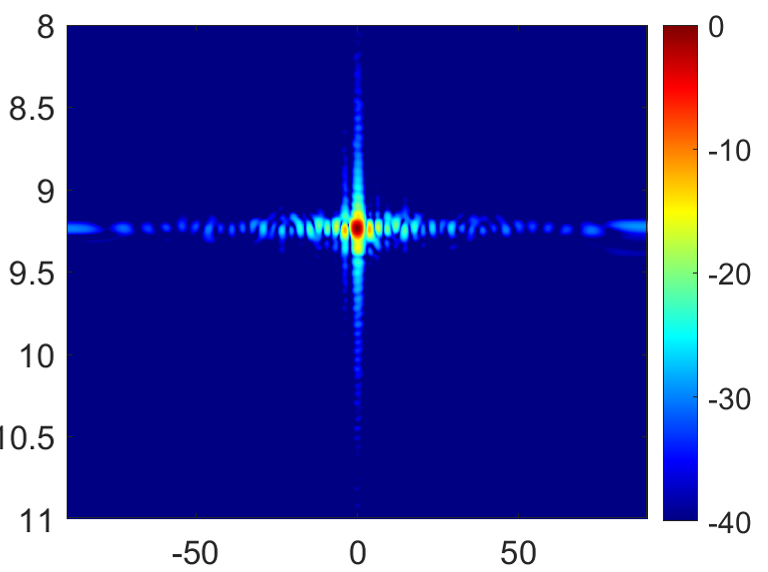}};

 \node[left=of img2, node distance=0cm, rotate = 90, xshift=0.6cm, yshift=-0.9cm,font=\color{black}] {{Range [m]}};
  \node[below=of img2, node distance=0cm, xshift=0cm, yshift=1.1cm,font=\color{black}] {{$\Theta ^ o$}};
 \node[below=of img2, node distance=0cm, xshift=0cm, yshift=0.8cm,font=\color{black}] {{(b)}};
\end{tikzpicture}

\caption{a) the 1D IPR along the elements of the virtual array using the data from the reference corner reflector, b) the 2D IPR of the array after phase calibration.}
\label{fig:IPR_no_cal_1D}
\end{figure}

In this section, we present the experimental results.
The virtual MIMO FMCW radar system that we utilize in this paper, is the 4-chip cascaded FMCW radar sensor TIDEP-01012 from TI with 12 TX and 16 RX antennas. The radar is operated in TDM mode which results in 86 non-overlapped virtual antenna elements. Therefore, it is considered as an 86-element FMCW radar system.
Fig.~\ref{fig:CASCADEBoard_1} illustrates the RF board which consists of 4 AWR1243 chips. Each AWR1243 chip has 3 TX and 4 RX antennas which after synchronization the entire board corresponds to a system of 12 TX and 16 RX antennas. Then, by exploiting the TDM technique we can achieve 192 elements virtually.
The effective time duration of the transmitted chirp signal is $40 \; \rm \mu s$. The center frequency of the radar has been set at $78.8 \; \rm GHz$. The effective bandwidth of the radar has been chosen to be $3.5997 \;\rm  GHz$ which theoretically\emph{} results in $3.7086\; \rm cm$ for the resolution limit in the range direction. Furthermore, the sampling frequency of the ADC is $8 \;\rm  MHz$.
The cut-off frequency for the high pass filters at the RF front-end, have been set to the smallest possible value.
The layout of the RF board has been depicted in Fig.~\ref{fig:layout} which clearly shows the distances between different elements that we need for the 3D imaging process.
There are two separate stepper motors which have been utilized to create 2D motion in the azimuth and elevation directions. The stepper motors are controlled by the Arduino DUE. One stepper motor is for generating linear motion in the horizontal direction and the other one is to create linear motion in the vertical direction. Consequently, we can create a 2D aperture. The Arduino DUE has been programmed using C++ programming language. The 4-chip cascaded radar board is controlled through the LUA commands provided by TI and are run in MATLAB environment. The stepper motors are controlled through a Python script with the help of the Arduino. A schematic of how different parts are connected has been presented in Fig.~\ref{fig:schematic}.
After the data is collected, it is stored in a flash memory on the DSP board which can be later transferred to the laptop through Ethernet connection.
When the data is stored in the laptop, the algorithm is applied and the image is generated subsequently. The algorithm for the image reconstruction has been written in MATLAB as well as Python.
Prior to discussing the image reconstruction procedure, we first perform single channel performance analysis for the radar system. Fig.~\ref{fig:spectrogram}-(a) illustrates the spectrogram of the beat signal for TX1-RX1 pair and for a $20\;\rm dbsm$ corner reflector which is located at the bore-sight of the system.  Fig.~\ref{fig:spectrogram}-(a) shows the effect of the non-linearity for the transmitted chirp signal.
In fact, the spectrogram should have been a straight line along the time axis. In Fig.~\ref{fig:spectrogram}-(b), we have presented the spectrogram for the same signal after compensating for the chirp non-linearity \cite{Meta_1, Meta_2}. As can be seen from Fig.~\ref{fig:spectrogram}-(b), the energy of the signal is distributed along a straight line horizontally over the entire chirp time.
Chirp non-linearity causes image de-focusing and blurring \cite{Meta_1, Meta_2}.
It should be noted that, the chirp non-linearity is different from channel to channel. Therefore, the chirp non-linearity compensation should be performed for each channel separately.
From Fig.~\ref{fig:spectrogram}, it is also obvious that the output power of the system is fluctuating over the entire bandwidth.
In Fig.~\ref{fig:power_spectrum}, the power spectrum for the same beat signal presented in Fig.~\ref{fig:spectrogram}, has been depicted which shows the power fluctuation with respect to frequency. The result is prior to performing any adjustment for the chirp non-linearity effect. However, the power fluctuation pattern does not change with the non-linearity compensation procedure. Similar to the chirp non-linearity, the power fluctuation is different for different channels.
From Fig.~\ref{fig:power_spectrum}-(a), we can see that there is a deep null around $80 \; \rm GHz$ which makes this portion of the chirp practically ineffective for the radar imaging process. In other words, the effective length of the chirp is considered to be shorter which reduces the range resolution limit of the system. 
After presenting the performance analysis for the single channel, we then focus on the array calibration concept.
Before applying the algorithm to the experimental data and performing image reconstruction, we address the process of the virtual MIMO array calibration.
The vertical direction is covered by the virtual MIMO array which consists of $86$ non-overlapped elements.
To calibrate the system, we use a $20\;\rm dbsm$ corner reflector.
The IPR of the array for the corner reflector, has been shown in Fig.~\ref{fig:power_spectrum}-(b).
As can be seen from Fig.~\ref{fig:power_spectrum}-(b), there is a strong imbalance for the side-lobe levels along the array elements. This implies the necessity for precise array calibration.
To obtain a better understanding of how to calibrate the array, we have focused on the IPR of the virtual MIMO array separately. In Fig.~\ref{fig:IPR_no_cal_1D}-(a), the 1D IPR of the virtual MIMO array has been shown, which is the horizontal section of the 2D IPR shown in Fig.~\ref{fig:power_spectrum}-(b) at the specific range cell, where the energy of the target is located along the array elements.
There are several different effects that can cause error. We consider the $l^{\rm {th}}$ channel and express the beat signal corresponding to a point reflector located at distance $R$ from the imagery system as
\begin{align}
\label{beat_signal_error_0}
s_l(t,R) = g_l\; p(t)\; e^{\left(\displaystyle -j 2 \pi (f_0 + \beta t) \frac{2(R+R_l)}{c}\right)},
\end{align}
where $g_l$ and $R_l$ are the gain and the error induced terms for the $l^{\rm {th}}$ channel, respectively.
Based on (\ref{beat_signal_error_0}), the error signal can then be expressed as
\begin{align}
\label{beat_signal_error}
\bar{s}_l(t) = g_l\; e^{\left(\displaystyle -j 2 \pi (f_0 + \beta t) \frac{2R_l}{c}\right)}.
\end{align}
We ignore other sources of error such as in-phase and quadrature imbalance and only focus on the three terms presented in (\ref{beat_signal_error}). The first error term is related to the gain of the channel, $g_l$. The gain varies from channel to channel. The second error term, which is the most important one, is the phase of $\bar{s}_l(t)$. Finally, the third part of the error is the frequency of the error signal $\bar{s}_l(t)$.
Upon taking Fourier transform from (\ref{beat_signal_error}), with the purpose of localizing the energy of the signal in the range direction, the frequency of $\bar{s}_l(t)$ will contribute to the range cell migration (RCM) phenomenon \cite{Cumming}. It spreads the energy of the signal over range cells other than the specific range cell that the energy of the target is supposed to be concentrated.
We can compensate for the effect of the frequency of $\bar{s}_l(t)$ by performing range alignment \cite{ChenISAR}.
If the point target is located at the bore-sight and in the far field zone, the phase of $\bar{s}_l(t)$ should be the same for all the channels. Therefore, any differences can be calculated and then their complex conjugate be applied to the raw data along different channels to compensate for the phase error term.
We can, further, compensate for the gain differences between different channels to complete the calibration procedure.
After performing phase calibration using the $20 \; \rm dbsm$ corner reflector, we obtain the result which has been shown in Fig.~\ref{fig:IPR_no_cal_1D}-(b).
From Fig.~\ref{fig:IPR_no_cal_1D}-(b), it is clear that the phase calibration has improved the IPR of the system considerably. In fact, the symmetry of the side-lobes are more pronounced compared to the situation presented in Fig.~\ref{fig:IPR_no_cal_1D}-(a). However, from Fig.~\ref{fig:IPR_no_cal_1D}-(b), it is also clear that the IPR is far from ideal.
The phase calibration has only taken into consideration the error corresponding to the phase of the signal given in (\ref{beat_signal_error}). The error corresponding to the frequency of $\bar{s}_l(t)$, given in (\ref{beat_signal_error}), is still to be removed. As we mentioned before, the frequency of $\bar{s}_l(t)$ contributes to the RCM phenomenon.
To see this effect, we have shown the range compressed image in Fig.~\ref{fig:range_compressed_rcmc}-(a), which is one step prior to the final image formation.
\begin{figure}[htb]
\begin{tikzpicture}[yshift=0.00001cm][font=\small]
\node(img1) {\includegraphics[height=2cm,width=3.5cm]{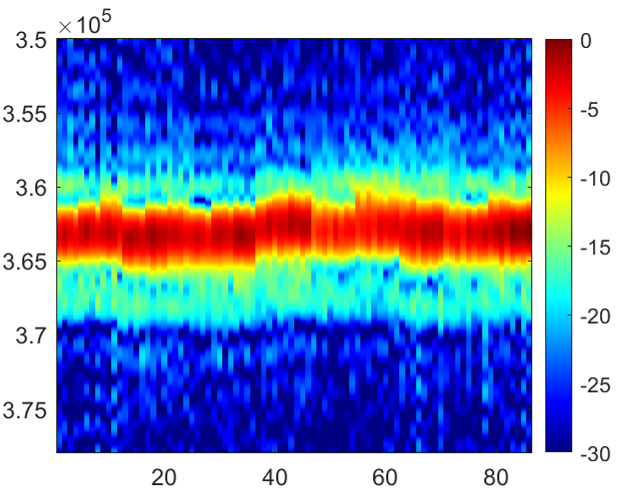}};

 \node[left=of img1, node distance=0cm, rotate = 90, xshift=1.2cm, yshift=-0.9cm,font=\color{black}] {{Range Samples}};
  \node[below=of img1, node distance=0cm, xshift=0cm, yshift=1.1cm,font=\color{black}] {{Element No.}};
 \node[below=of img1, node distance=0cm, xshift=0cm, yshift=0.7cm,font=\color{black}] {{(a)}};
\hspace{4.4cm}
\node(img2) {\includegraphics[height=2cm,width=3.5cm]{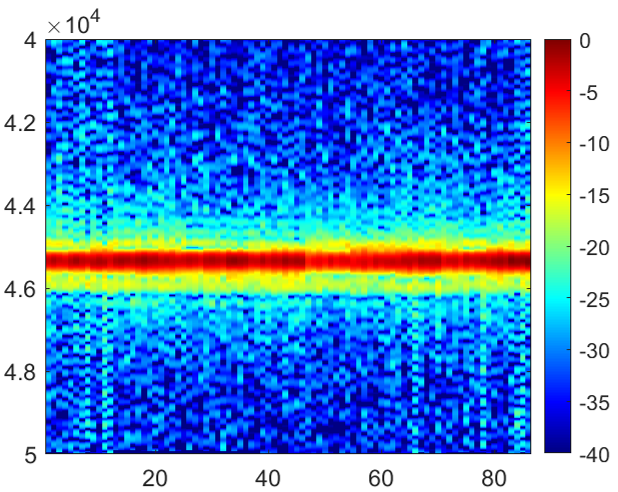}};

 \node[left=of img2, node distance=0cm, rotate = 90, xshift=1.2cm, yshift=-0.9cm,font=\color{black}] {{Range Samples}};
  \node[below=of img2, node distance=0cm, xshift=0cm, yshift=1.1cm,font=\color{black}] {{Element No.}};
 \node[below=of img2, node distance=0cm, xshift=0cm, yshift=0.7cm,font=\color{black}] {{(b)}};

\end{tikzpicture}
\caption{The range compressed image, a) before compensating for the RCM effect, b) after performing range alignment.}
\label{fig:range_compressed_rcmc}
\end{figure}
To concentrate the energy of the target in one range cell and therefore compensate for the effect of the RCM, we perform range alignment \cite{ChenISAR}. The result of the range aligned image has been illustrated in Fig.~\ref{fig:range_compressed_rcmc}-(b).
The result for the IPR of the system, after phase calibration and range alignment, has been shown in Fig.~\ref{fig:IPR_with_cal_2D_range_alignment}.
\begin{figure}[htb]
\centering
\begin{tikzpicture}[yshift=0.05cm][font=\small]
  \node (img1)  {\includegraphics[height=2.5cm,width=4cm]{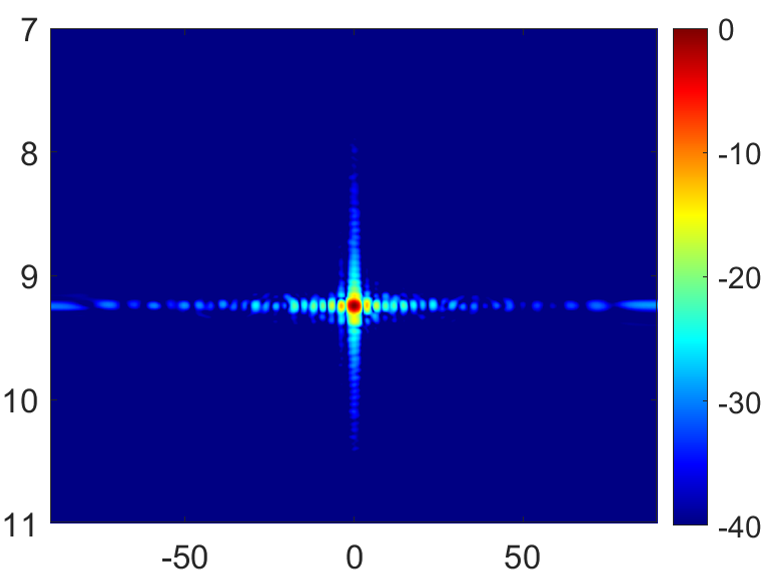}};
 \node[left=of img1, node distance=0cm, rotate = 90, xshift=1cm, yshift=-0.9cm,font=\color{black}] {{Range [m]}};
  \node[below=of img1, node distance=0cm, xshift=-0.1cm, yshift=1.1cm,font=\color{black}] {{$\Theta ^ o$}};
\end{tikzpicture}
\caption{The IPR of the array after phase calibration and range alignment.}
\label{fig:IPR_with_cal_2D_range_alignment}
\end{figure}
Furthermore, Fig.~\ref{fig:IPR_with_cal_1D}-(a) illustrates the horizontal section of the 2D IPR in the direction parallel to the array elements at the location of the main-lobe. Basically, Fig.~\ref{fig:IPR_with_cal_1D}-(a) depicts the 1D IPR along the array elements.
\begin{figure}[htb]
\begin{tikzpicture}[yshift=0.00001cm][font=\tiny]
\node(img1) {\includegraphics[height=2cm,width=3.5cm]{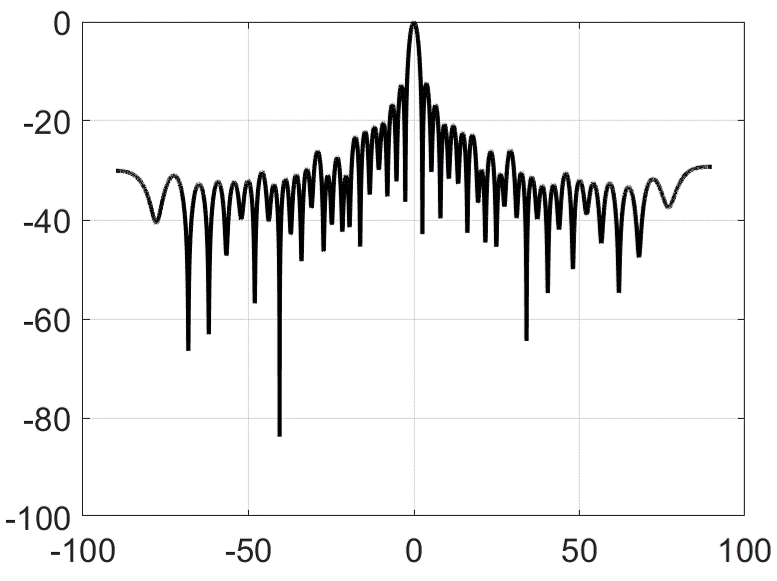}};

 \node[left=of img1, node distance=0cm, rotate = 90, xshift=1cm, yshift=-0.9cm,font=\color{black}] {{Normalized Power [dB]}};
  \node[below=of img1, node distance=0cm, xshift=0cm, yshift=1.1cm,font=\color{black}] {{$\Theta^o$}};
 \node[below=of img1, node distance=0cm, xshift=0cm, yshift=0.8cm,font=\color{black}] {{(a)}};
\hspace{4.4cm}
\node(img2) {\includegraphics[height=2cm,width=3.5cm]{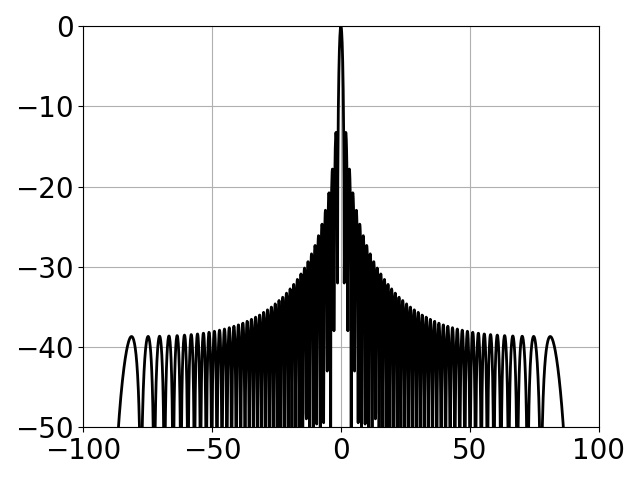}};

\node[left=of img2, node distance=0cm, rotate = 90, xshift=1cm, yshift=-0.9cm,font=\color{black}] {{Normalized Power [dB]}};
  \node[below=of img2, node distance=0cm, xshift=0cm, yshift=1.1cm,font=\color{black}] {{$\Theta^o$}};
 \node[below=of img2, node distance=0cm, xshift=0cm, yshift=0.8cm,font=\color{black}] {{(b)}};

\end{tikzpicture}
\caption{The 1D IPR along the virtual array elements, a) after phase calibration and range alignment, b) after phase calibration, range alignment and applying the optimization problem given in (\ref{opt}).}
\label{fig:IPR_with_cal_1D}
\end{figure}
However, based on Fig.~\ref{fig:IPR_with_cal_1D}-(a), we can see that the IPR is not equal to the ideal IPR for a linear array with $86$ transceivers.
Therefore, we apply the following optimization problem to further compensate for any remaining phase errors.
\begin{equation}
\label{opt}
\begin{array}{rrclcl}
\displaystyle \min_{\bw} & \multicolumn{3}{l}{\| (\bw \bigodot \bs)\bA - \mathbf{1}^{T}\bA \|_2},\\
&\bw \in \mathcal{C},   \\
\end{array}
\end{equation}
where $\bw$ is a 1D complex vector and $\bs$ is the signal presented in Fig.~\ref{fig:IPR_with_cal_1D}. Moreover, $\bigodot$ is the element-wise multiplication operator and matrix $\bA$ represents the array manifold. Finally, $\mathbf{1}$ is a column-wise vector with all elements equal to 1.
The optimization problem presented in (\ref{opt}) is a convex optimization problem \cite{Boyd} and can be solved using available packages such as $\rm cvx$ in MATLAB or $\rm cvxpy$ in Python.
After applying the optimization problem to the signal shown in Fig.~\ref{fig:IPR_with_cal_1D}-(a), we achieve the result illustrated in Fig.~\ref{fig:IPR_with_cal_1D}-(b) for the 1D IPR of the system.
Before performing the image reconstruction process based on the cascaded-chip radar system, we manage to obtain an mmWave image using a single TX-RX pair, which basically means we are conducting single channel-based imaging. We are utilizing a single chip AWR1243 FMCW radar from TI to create a 3D image from the sample under test. To create the image we use the data from a single TX-RX pair \cite{Shahrokh_SAR}. In fact, we are bench-marking the performance of the virtual MIMO-SAR FMCW radar against the single channel system.
The sample under test, which is a metallic object, has been shown in Fig.~\ref{fig:test_sample}.
The distance between the sample under test and the aperture is $z_d = 30 \; \rm  cm$.
\begin{figure}[htb]
\centering
\begin{tikzpicture}[yshift=0.00001cm][font=\small]
\node(img1) {\includegraphics[height=3cm,width=5cm]{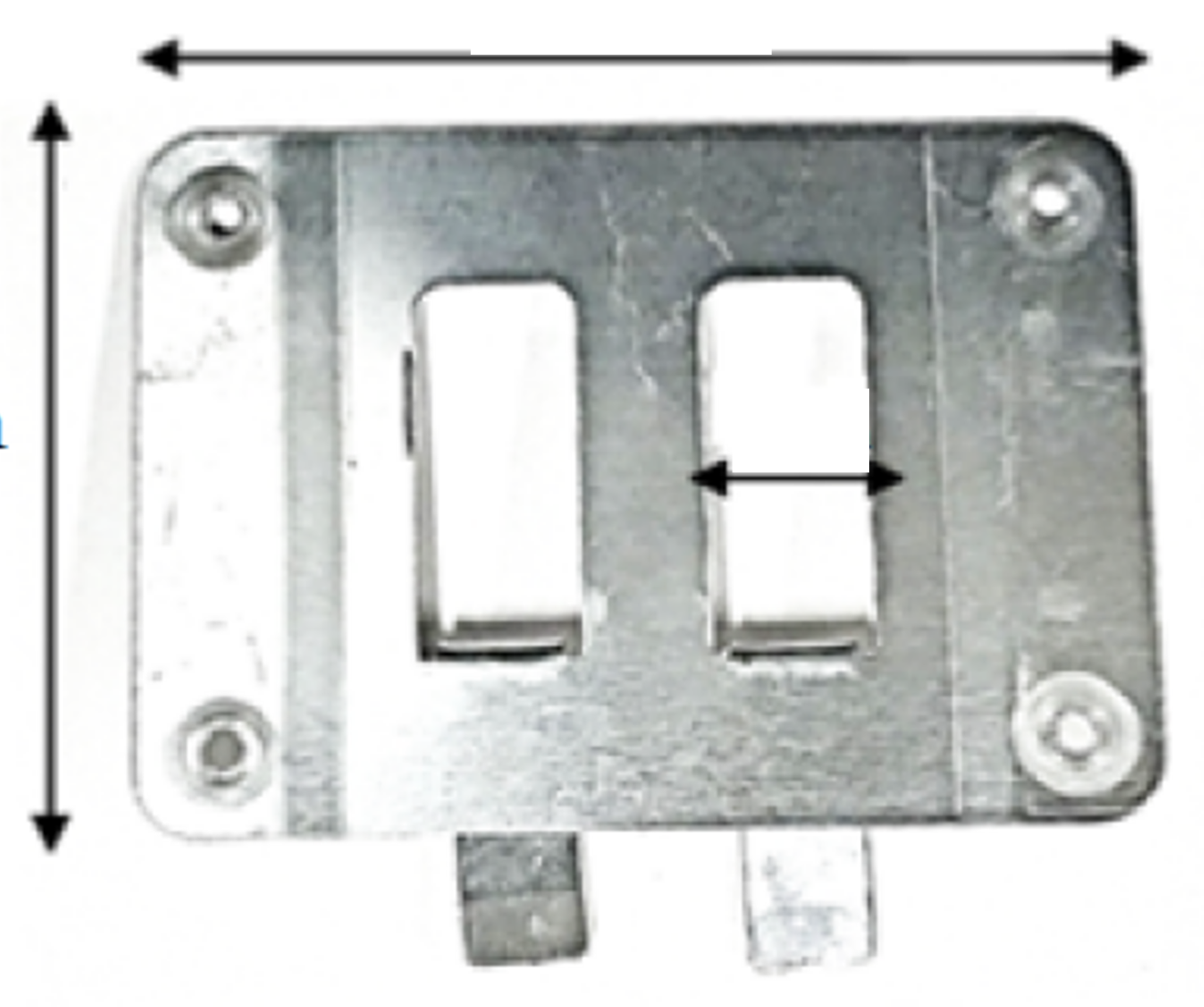}};
 \node[left=of img1, node distance=0cm, xshift=1.2cm, yshift=0.1cm,font=\color{black}] {{7 cm}};
 \node[left=of img1, node distance=0cm, xshift=4.2cm, yshift=1.6cm,font=\color{black}] {{10.5 cm}};
\node[left=of img1, node distance=0cm, xshift=4.9cm, yshift=0.3cm,font=\color{black}] {{2 cm}};
\end{tikzpicture}
\caption{The sample under test.}
\label{fig:test_sample}
\end{figure}
In Fig.~\ref{fig:img_single_channel}-(a), the image acquired using the data gathered from a single channel radar system, has been shown.
The 2D aperture has been created synthetically in the azimuth and elevation directions. The time duration for the data collection is approximately 7 minutes which can be reduced considerably by exploiting the virtual MIMO array in the vertical direction.
\begin{figure}[htb]
\centering
\begin{tikzpicture}[yshift=0.00001cm][font=\small]
  \node (img1)  {\includegraphics[height=4.5cm,width=6cm]{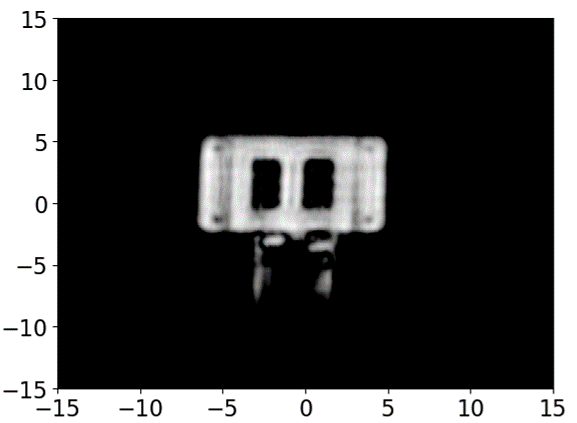}};
   \node[left=of img1, node distance=0cm, rotate = 90, xshift=0.8cm, yshift=-0.9cm,font=\color{black}] {{Y [cm]}};
  \node[below=of img1, node distance=0cm, xshift=0cm, yshift=1.1cm,font=\color{black}] {{X [cm]}};
 \node[below=of img1, node distance=0cm, xshift=0cm, yshift=0.7cm,font=\color{black}] {{(a)}};
\end{tikzpicture}

\begin{tikzpicture}[yshift=0.05cm][font=\small]
  \node (img2)  {\includegraphics[height=5.5cm,width=8cm]{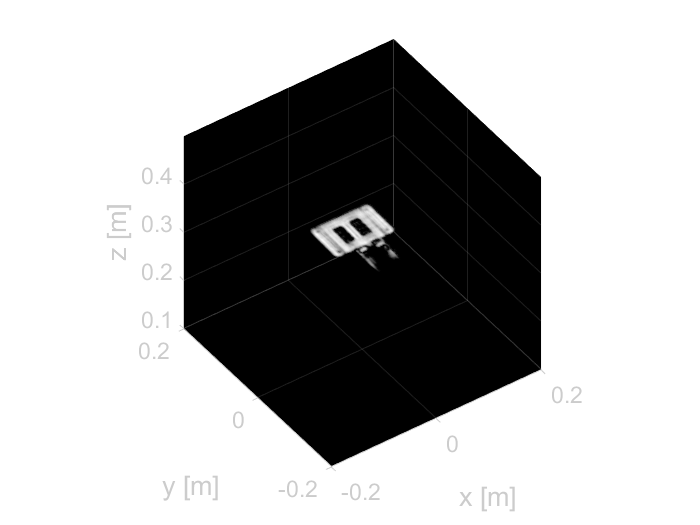}};
\node[below=of img2, node distance=0cm, xshift=0cm, yshift=1.2cm,font=\color{black}] {{(b)}};
\end{tikzpicture}

\caption{a) the reconstructed image from the sample under test based on a single TX-RX pair from the AWR1243 FMCW radar chip, b) the 3D view of the reconstructed image.}
\label{fig:img_single_channel}
\end{figure}
Moreover, Fig.~\ref{fig:img_single_channel}-(b) shows the 3D view of the reconstructed image.
After performing array calibration, we can exploit the virtual MIMO array to cover the vertical direction of the 2D aperture which we are at the point of creating for the 3D imaging process. The sample under test can be covered by the 86 non-overlapped elements of the virtual MIMO array, therefore, there is no requirement to move the virtual MIMO array in the vertical direction. Consequently, in the vertical direction we have the virtual MIMO array and in the horizontal direction the aperture is created synthetically and hence the name virtual MIMO-SAR imaging.
\begin{figure}
\centering
\includegraphics[height=4cm,width=6cm]{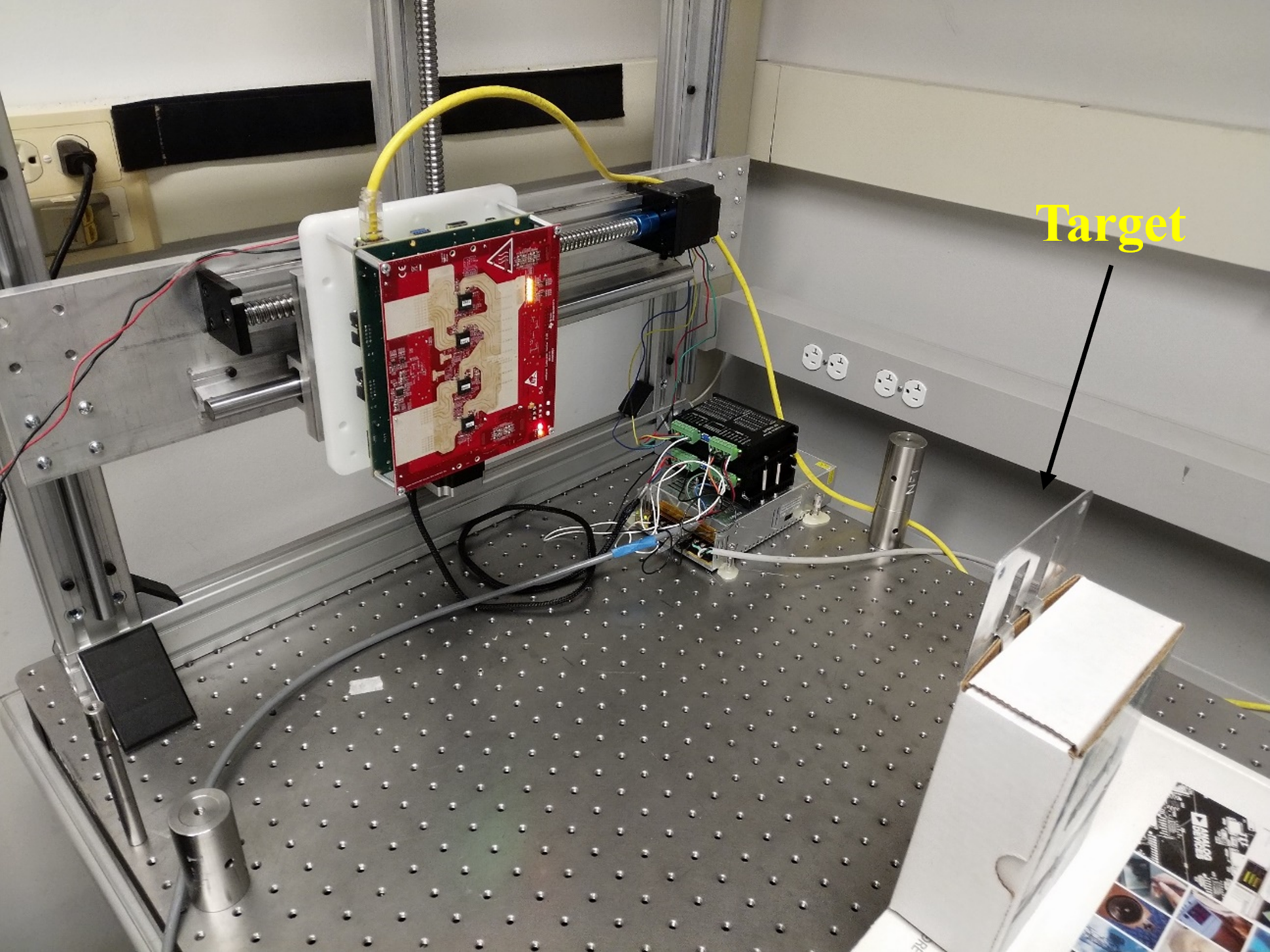}
\caption{ The experimental set-up.
\label{fig:TestSample}}
\end{figure}

\begin{figure}[htb]
\centering
\begin{tikzpicture}[yshift=0.00001cm][font=\small]
  \node (img1)  {\includegraphics[height=4.5cm,width=6cm]{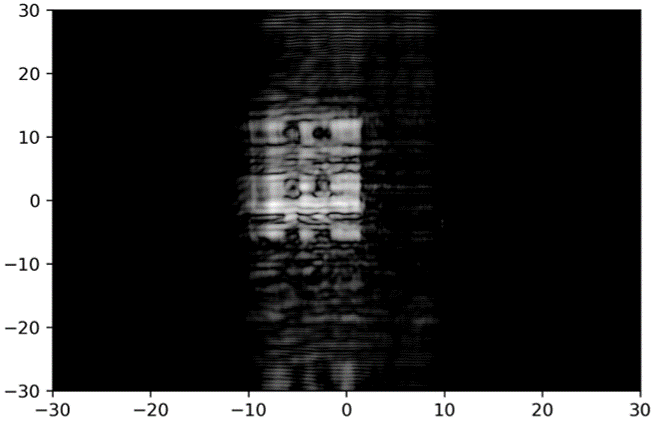}};
   \node[left=of img1, node distance=0cm, rotate = 90, xshift=0.8cm, yshift=-0.9cm,font=\color{black}] {{Y [cm]}};
  \node[below=of img1, node distance=0cm, xshift=0cm, yshift=1.1cm,font=\color{black}] {{X [cm]}};
 \node[below=of img1, node distance=0cm, xshift=0cm, yshift=0.7cm,font=\color{black}] {{(a)}};
\end{tikzpicture}

\begin{tikzpicture}[yshift=0.00001cm][font=\small]
  \node (img1)  {\includegraphics[height=4.5cm,width=6cm]{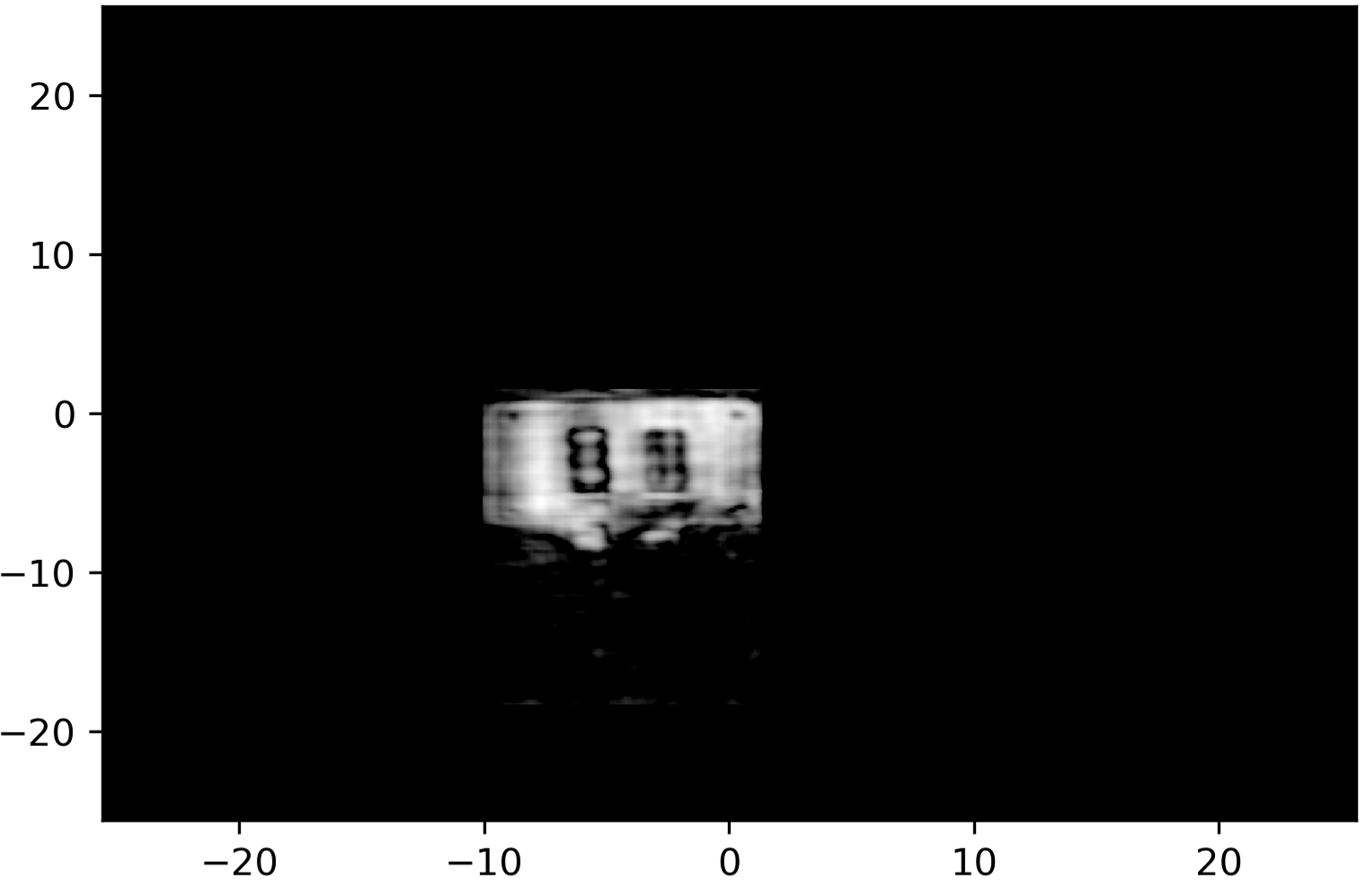}};
   \node[left=of img1, node distance=0cm, rotate = 90, xshift=0.8cm, yshift=-0.9cm,font=\color{black}] {{Y [cm]}};
  \node[below=of img1, node distance=0cm, xshift=0cm, yshift=1.1cm,font=\color{black}] {{X [cm]}};
 \node[below=of img1, node distance=0cm, xshift=0cm, yshift=0.7cm,font=\color{black}] {{(b)}};
\end{tikzpicture}

\caption{a) the resulting image without array calibration, b) the reconstructed image after performing calibration on the virtual MIMO array.}
\label{fig:final_image}
\end{figure}
In Fig.~\ref{fig:TestSample}, the experimental set-up has been shown which illustrates the cascaded-chip radar, the 2D aperture, and the sample under test.
In Fig.~\ref{fig:final_image}-(a) the reconstructed image, for the un-calibrated data, has been presented. Finally, Fig.~\ref{fig:final_image}-(b) illustrates the reconstructed image after the calibration procedure has been applied to the raw data.
The image has been reconstructed based on (\ref{RMA_Algorithm_f}).
Fig.~\ref{fig:final_image}-(a) shows clearly the effect of the errors associated with the MIMO array in the reconstructed image along the vertical direction which upon applying the calibration procedure, the errors are removed and the reconstructed image is focused.
The speed of the stepper motor has been set to $5 \; \rm cm/s$. Only a single position in the elevation direction has been used for the data collection.
The number of elements in the elevation direction is $86$. It has been achieved based on the virtual MIMO system which allows to increase the number of elements virtually by exploiting the TDM method and sparsity of the array. In the azimuth direction, we have created the effect of an array with $197$ elements by utilizing  the SAR concept.
The entire data collection has taken $4 \;\rm s$.

From Fig.~\ref{fig:final_image}-(b), we can see that the final result still suffers from a certain amount of remaining errors. This can be attributed to the fact that we did not perform a precise measurement on the location of the corner reflector which we used for array calibration. This can contribute to the remaining errors in different ways.

To begin with, we can mention the multi-static to mono-static transformation which was presented by the phase term given in (\ref{beat_signal_continuum_3}). The phase term $\Phi (\Delta_x, \Delta_y, z_d)$ expressed in (\ref{beat_signal_continuum_3}) is range dependent and when we apply the multi-static to mono-static transformation to the calibration data since we have not measured the location of the corner reflector precisely, therefore, we should anticipate phase errors generated by the transformation.

Another source of the phase error can also be attributed to the imprecise measurement for the calibration data. The assumption behind the calibration is based on having a corner reflector at bore-sight. In reality, however, the corner reflector might not be located at bore-sight precisely which results in certain amount of phase errors which subsequently blurs the reconstructed image.

Moreover, when we are utilizing the entire 86-non overlapped antennas, we will be estimating the multi-static radar cross section of the target. The multi-static radar cross section can be different from the cross section that we estimate when we utilize one specific pair of  TX-RX.

Furthermore, the RF board of the 4-chip-cascaded radar, that we have utilized for our experiment in this paper, has not been completely optimized. From the 2D IPR of the system, it is clear that there is a large amount of coupling between the antenna elements. The feed-lines as well as the antennas should be re-designed in a way to mitigate the effect of the coupling between the elements.  The effect of the coupling between the elements on the TX side is to lower the efficiency of the system since a portion of the transmitted energy is absorbed by the adjacent elements. On the RX side, however, the coupling between the elements affects the beam-forming process. Basically, the signals received by different RXs become correlated and this in turn will hinder the beam-forming process. 

It should be noted that the horizontal direction is also prone to errors related to the non-uniformity of the linear motion along the $x$ direction. The non-uniform motion in the $x$ direction can be translated into non-uniform sampling phenomenon and can be addressed through interpolation-based techniques \cite{Soumekh}. However, there might be vibrations in the $y$ and $z$ directions as well which are further cause of image de-focusing. These undesired effects can be compensated by exploiting techniques such as array calibration, which we have used to calibrate the MIMO array, or by applying motion compensation methods \cite{ChenISAR, Soumekh}. Another effective technique, is auto-focus approach which is used extensively in SAR and ISAR image formation to further focus the reconstructed image \cite{ChenISAR}.

\begin{figure}[htb]
\centering
\begin{tikzpicture}[yshift=0.00001cm][font=\small]
  \node (img1)  {\includegraphics[height=4.5cm,width=6cm]{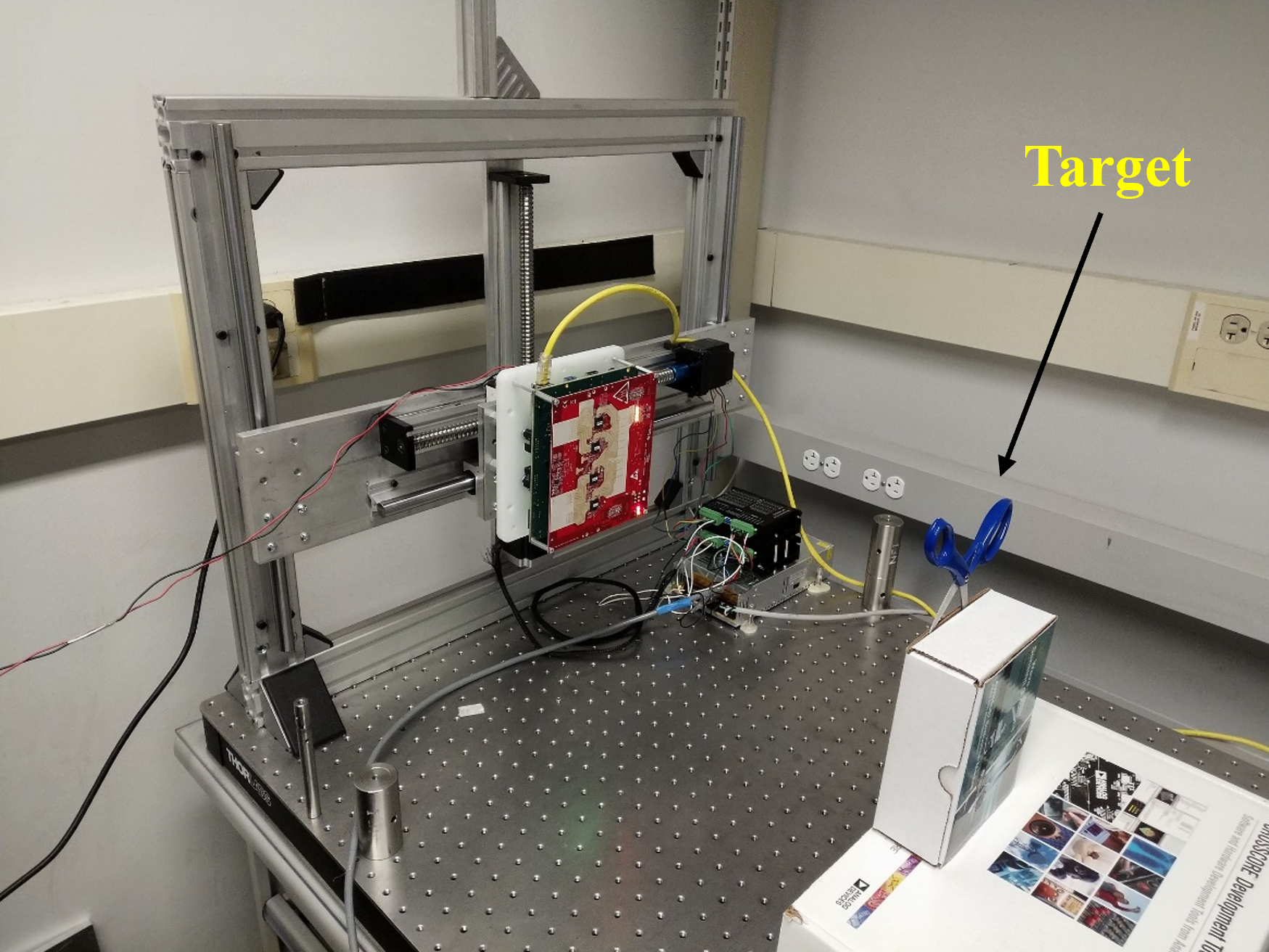}};
  \node[below=of img1, node distance=0cm, xshift=0cm, yshift=0.8cm,font=\color{black}] {{(a)}};
\end{tikzpicture}

\begin{tikzpicture}[yshift=0.00001cm][font=\small]
  \node (img1)  {\includegraphics[height=4.5cm,width=6cm]{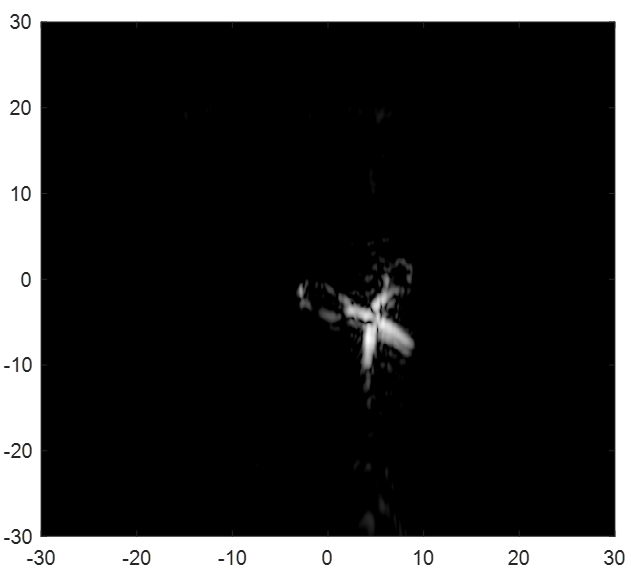}};
   \node[left=of img1, node distance=0cm, rotate = 90, xshift=0.8cm, yshift=-0.9cm,font=\color{black}] {{Y [cm]}};
  \node[below=of img1, node distance=0cm, xshift=0cm, yshift=1.1cm,font=\color{black}] {{X [cm]}};
 \node[below=of img1, node distance=0cm, xshift=0cm, yshift=0.7cm,font=\color{black}] {{(b)}};
\end{tikzpicture}

\caption{a) the experimental set-up for a pair of scissors in front of the imagery system, b) the reconstructed image.}
\label{fig:final_image_2}
\end{figure}

Finally, we have conducted another test experiment in which a pair of scissors have been used. Fig.~\ref{fig:final_image_2}-(a) shows the experimental set-up. In Fig.~\ref{fig:final_image_2}-(b), we have presented the reconstructed image after performing array calibration. The image is based on all 86 non-overlapped elements of the virtual MIMO system. 

It should be noted that, the calibration process of the MIMO array for high resolution imaging, similar to the work presented in this paper, is challenging. The main focus of our future work is to improve the calibration technique, which has been presented in this paper, or to discover a completely new method for the MIMO array calibration.   
\section{Conclusion}
In this paper, we presented 3D near-field high resolution mmWave radar imaging technique based on a combination of virtual MIMO array and SAR. We described the model as well as the image reconstruction method at length and at the end applied the algorithm to the experimental data gathered from a virtual MIMO array FMCW radar operating at $77\;\rm GHz$ and presented the final results. Combination of virtual MIMO array system and SAR concept offers a compact and cost-effective imagery system with lower complexity and higher speed which can be used to produce high resolution 3D radar images.

\bibliographystyle{IEEEtran}
\bibliography{Biblio}

\begin{IEEEbiography}[{\includegraphics[width=2.9cm,height=3.3cm,clip,keepaspectratio]{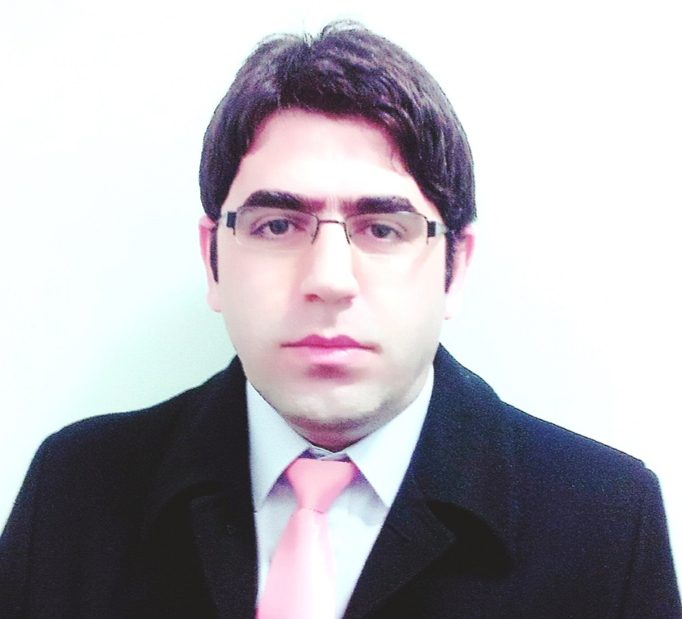}}]
{Shahrokh Hamidi}
was born in Iran, in 1983. He received his B.Sc., M.Sc., and Ph.D. degrees all in Electrical and Computer Engineering. He is with the faculty of Electrical and Computer Engineering at the University of Waterloo, Waterloo, Ontario, Canada. His current research areas include statistical signal processing, mmWave imaging, Terahertz imaging, image processing, system design,  multi-target tracking, wireless communication, machine learning, optimization, and array processing.
\end{IEEEbiography}

\end{document}